\documentstyle[aps,pra,subfigure,epsfig,amsmath,amssymb,float,12pt]{revtex}
\newcommand{\bsym}[1]{\hbox{\rm \boldmath{$#1$}}}
  
\begin{document}

\draft

%\title{Transport behaviour in mixtures of Bose-Einstein condensates 
%with thermal atoms: tests of an explicit time-marching algorithm
%combined with a direct simulation Monte Carlo method}

\title{Explicit finite-difference and direct-simulation-MonteCarlo method
for the dynamics of mixed Bose-condensate and cold-atom clouds} 

\author{P. Vignolo$^1$, M.~L. Chiofalo$^1$, S. Succi$^2$,
and M.~P. Tosi$^1$}

\address{$^1$
INFM and
Classe di Scienze, Scuola Normale Superiore, Piazza dei Cavalieri 7,
I-56126 Pisa, Italy\\
$^2$Istituto Applicazioni Calcolo ``M.~Picone'',
Viale del Policlinico 137, I-00161 Roma, Italy\\
}
\maketitle
\vspace{0.5cm}

\begin{abstract}
We present a new numerical method for studying the dynamics of
quantum fluids composed of a Bose-Einstein condensate and a cloud 
of bosonic or fermionic atoms in a mean-field approximation.
It combines an explicit time-marching algorithm, previously developed
for Bose-Einstein condensates in a harmonic or optical-lattice potential, 
with a particle-in-cell MonteCarlo approach to the equation of motion for 
the one-body Wigner distribution function in the cold-atom cloud. The method 
is tested against known analytical results on the free expansion of a fermion
cloud from a cylindrical harmonic trap and is validated by examining how
the expansion of the fermionic cloud is affected by the simultaneous expansion 
of a condensate. We then present wholly original calculations on
a condensate and a thermal cloud inside a harmonic well
and a superposed optical lattice, by addressing the free expansion of
the two components and their oscillations under an applied harmonic force.
These results are discussed in the light of relevant theories and experiments.
\end{abstract}

\vspace{0.5cm}

\section{Introduction}  
The physics of ultracold atomic vapours under magnetic or optical
confinement has been a continuing and ever expanding focus of interest
since the realization of Bose-Einstein condensation~\cite{BECexp}.
Following characterizations of the basic thermodynamic and dynamical
properties of condensates~\cite{JILAth}, a number of 
experiments have been addressed to investigations of their phase 
coherence and superfluidity~\cite{Superfluid}, to the study of non-linear
effects and special spectroscopic features ~\cite{4WM}, and to the 
observation of vortices~\cite{Vortices}. Parallel efforts are being
made in the study of gases of fermionic atoms~\cite{FERexp} and of
boson-fermion mixtures~\cite{sette}, with the ultimate aim of 
realizing novel superfluids.

Theoretical studies of the dynamics of these systems, involving the
analytical
solution of approximate models have very often been 
successful in explaining or predicting such novel
phenomena. However, the interplay of different species
or the thermal fluctuations of a condensate are not easily 
handled by analytical methods.
The validity of the approximations may be limited to 
the extreme collisionless or hydrodynamic regimes, 
the confinement of
the sample being treated within a local-density approach. 
Thus, while the equations governing these
dilute systems remain simple, their numerical solution can be 
helpful for investigating more complex dynamical problems where 
an intermediate regime is met or a cold-atom cloud accompanies
the condensate.

The atomic interactions in a highly dilute Bose gas at very low
temperatures, as is relevant for current experiments, are described
by a contact pseudopotential accounting for $s$-wave scattering in 
binary atom-atom elastic collisions. A condensate is then treated
within a mean-field approach by solving the stationary
or the time-dependent Gross-Pitaevskii equation (GPE).
Several types of numerical approaches
have been developed: eigenvalue solvers\cite{EIGEN}, 
variational solvers\cite{VAR}, or explicit solvers \cite{PREGS}
for the ground state, and implicit \cite{IMPLICIT} or explicit 
\cite{PRE} time-marching schemes for the dynamics.

Methods for studying the dynamics of 
an isolated cloud of ultracold (bosonic or fermionic) 
atoms are also well developed. One needs
to solve the Vlasov-Landau equation of motion (VLE) for the Wigner
distribution function.
Various numerical techniques have been used for this purpose, 
which are based either on an ergodic assumption~\cite{Murray} or on the 
inclusion of statistical noise~\cite{IACOPOALICE}, or else use a direct 
simulation MonteCarlo (DSMC) approach borrowed from the methods of 
molecular dynamics~\cite{Bird,Smerzi,Wu}.

The general theoretical background is provided by the book of
Kadanoff and Baym \cite{KB}. These authors developed a Green's function 
approach to
transport phenomena, which extends the Boltzmann equation to strongly 
interacting 
quantum fluids and allows for progressively improved
self-consistent approximations. 
This formalism was extended to a homogeneous
Bose-condensed gas at finite temperature by Kane and Kadanoff~\cite{KK},
within a Beliaev
approximation including interactions up to second order. 
More recently, these methods have been 
adapted to the theory of transport phenomena in a confined Bose gas 
within the Hartree-Fock-Bogoliubov approximation \cite{Griffin,ZGN,Rimurray},
dealing with a Bose-Einstein condensate accompanied by its thermal cloud.
Jackson and Adams \cite{Adams} have proposed to combine
the GPE with a quantum version of the DSMC to numerically evaluate
the dynamics of such a fluid. Numerical studies based on a generalized GPE
combined with a semiclassical kinetic equation, and including collisions 
between the condensate and the thermal cloud, are also becoming avaible
for dynamical properties of a trapped Bose-condensed gas at finite temperature
~\cite{Zaremba}.

In the present work we proceed along the path traced by Jackson and 
Adams~\cite{Adams}. We propose a different approach to the solution
of the GPE and a different method for
preparing the initial equilibrium state, which would be immediately
applicable to a multi-component cold-atom cloud.
The method is applied to two classes of dynamical problems: the first
concerns the ballistic expansion of a fermion cloud and the role played
by the presence of a condensate, while the second concerns a Bose-condensed
gas in a periodic optical-lattice potential at finite temperature.

After introducing in Sec.~\ref{Model} 
the model for both the equilibrium state and the
dynamical evolution of the fluid,  
we describe in Sec.~\ref{Method} 
the numerical methods that we have used to consistently
solve the GPE for the condensate and the VLE for the Wigner distribution
function of the cloud. 
The procedure followed in the actual computations is also outlined
in Sec.~\ref{Method}, with due emphasis on the preparation  
of the initial equilibrium state. The physical applications that
we have carried out are presented and discussed in Sec.~\ref{Results}.
A discussion of computational aspects in Sec.~\ref{remark} and some
final remarks in Sec.~\ref{conclusions} conclude the paper.

%SONO QUI

\section{The model: a review of the main concepts}\label{Model}
\subsection{Transport in a normal quantum fluid}

In the original formulation by
Kadanoff and Baym (KB), 
the problem of transport in a quantum fluids in the normal state
 is tackled by
deriving an analogue of the Boltzmann equation for the Wigner 
distribution function $f({\bf p},{\bf r},t)$ from the microscopic 
equations of motion for the non-equilibrium density matrix
$\rho({\bf r}_1,t_1;{\bf r}_{1'}t_{1'})\equiv
<\psi^{\dagger}_{\cal U}({\bf r}_1,t_1)\psi^{}_{\cal U}({\bf r}_{1'},t_{1'})>$, which is
defined through the particle creation and destruction operators 
$\psi_{\cal U}^{(\dagger)}$ in the 
presence of an external, slowly varying disturbance ${\cal U}({\bf r},t)$. 
Namely,
\begin{equation}
f({\bf p},{\bf r},t)=\int d{\bf x}\exp(-i{\bf p}\cdot
{\bf x})
<\psi^{\dagger}_{\cal U}({\bf r}+{\bf x}/2,0)
\psi^{}_{\cal U}({\bf r}-{\bf x}/2,t)>\; 
\label{wigner}
\end{equation}
where 
${\bf r}=({\bf r}_1+{\bf r}_{1'})/2$ and
${\bf x}={\bf r}_1-{\bf r}_{1'}$ are the 
center-of-mass and the relative coordinate of the two particles.
The moments of the Wigner function 
yield observables such as  
the particle density 
$n({\bf r},t)=(2\pi)^{-3}\int d{\bf p} f({\bf p},{\bf r},t)$ and the
current density 
${\bf j}({\bf r},t)=(2\pi)^{-3}
\int d{\bf p}\, ({\bf p}/m)\, f({\bf p},{\bf r},t)$.

Contact with the Boltzmann transport equation is made by performing
gradient expansions.
As in the conventional Boltzmann-equation approach,  
the validity of the KB formulation is limited to slowly varying 
perturbations. On the other hand, the advantage of the KB formulation
is that higher-order correlations enter the equation of motion
for the density matrix
in an explicit manner, and therefore systematically improved approximations 
which are consistent with the conservation laws are accessible. 
Examples of such treatments of the correlation term are
the Hartree approximation and the Born-collision 
approximation. In the former case
the collisionless Boltzmann equation is recovered, while
in the latter the collisional Boltzmann equation 
is extended to non-dilute systems by including the effect
of the external potential on the motion of the particles between
collisions. 

\subsection{Extension to coupled condensate-noncondensate dynamics}
The extension of the KB treatment to
gases including a Bose-condensed
component has been made by 
Kane and Kadanoff \cite{KK} and further developed by Griffin and 
coworkers~\cite{Griffin,ZGN} and also by 
Wachter {\it et al.} \cite{Rimurray} through a different 
derivation. The presence of two components and 
the appearence
of off-diagonal elements in the density matrix (the so-called
anomalous densities) below the Bose-Einstein condensation temperature 
requires the introduction of 
three Wigner distribution functions:  
$f_c({\bf p},{\bf r},t)$
for the condensate component described by
$<\psi^{(\dagger)}_{\cal U}>$ and involving $|<\psi^{\dagger}_{\cal U}>|^2$,
${f_b}({\bf p},{\bf r},t)$
for the noncondensate described by the fluctuations operators
$\tilde{\psi}^{(\dagger)}_{\cal U}\equiv
\psi^{(\dagger)}_{\cal U}-<\psi^{(\dagger)}_{\cal U}>$ and involving
$<\tilde{\psi}^\dagger_{\cal U}\tilde{\psi}_{\cal U}>$, and
${f}_{m}({\bf p},{\bf r},t)$ for the anomalous part
involving $<\tilde{\psi}_{\cal U}\tilde{\psi}_{\cal U}>$ and its Hermitean
conjugate.

We thus have to deal with the density of condensate 
$n_c({\bf r},t)=(2\pi)^{-3}\int d{\bf p}\, 
f_c({\bf p},{\bf r},t)$, the density of noncondensate 
$n_b({\bf r},t)=(2\pi)^{-3}\int d{\bf p}\, {f_b}({\bf
p},{\bf r},t)$, and the anomalous density 
${m}({\bf r},t)=(2\pi)^{-3}\int d{\bf p}\, {f}_m({\bf
p},{\bf r},t)$. Analogous expressions holds
for the current densities. 

As to the consistency of the approximations with the conservation 
laws, the same general remarks 
as for normal systems apply. However, 
the appearence of the condensate 
introduces an additional principle of gauge invariance, leading
to the requirement that 
the excitation spectrum to be gapless \cite{HM}. 
It is well known \cite{Griffin,HM} that approximations capable of 
simultaneously accommodating the 
conservation laws and the gaplessness condition 
are hardly available, so that a choice has to 
be made depending on the specific conditions of density and 
temperature of the system. 

In the regime that we address in the present work 
the anomalous densities 
can be neglected, resulting in the gapless
Hartree-Fock-Bogoliubov-Popov approximation
\cite{Griffin}. Thus, the equation of motion for the condensate 
wavefunction $\Phi({\bf r},t)\equiv<\psi^{\dagger}_{\cal U}({\bf r},t)>$ is
\begin{equation}
i\hbar\frac{\partial\Phi({\bf r},t)}{\partial t}=
\left[-\frac{\hbar^2}{2m}\bsym{\nabla}^2+
V^{eff}_{c,b}({\bf r},t)\right]\Phi({\bf r},t)
\label{eqc}
\end{equation}
and is coupled to the collisionless Vlasov equation for the noncondensate 
Wigner function
${f}_b({\bf p},{\bf
r},t)$, 
\begin{equation}
\frac{\partial {f}_b({\bf p},{\bf r},t)}
{\partial t}+\frac{{\bf p}}{m}\cdot\bsym{\nabla_{\bf r}}
{f}_b({\bf p},{\bf r},t)-
\bsym{\nabla_{\bf r}}V^{eff}_{b}({\bf r},t)\cdot\bsym{\nabla}_{\bf
p}{f}_b({\bf p},{\bf r},t)=0\; .
\label{eqf}
\end{equation}
The mean-field potentials  in Eqs. (\ref{eqc}) and (\ref{eqf}) are
\begin{equation}
V^{eff}_{c,b}({\bf r},t)=V^{ext}_b({\bf r})+{\cal U}({\bf r},t)+
U_g[n_c({\bf r},t)+2{n}_b({\bf
r},t)]\; 
\label{veffab}
\end{equation}
and
\begin{equation}
V^{eff}_{b}({\bf r},t)=V^{ext}_b({\bf r})+{\cal U}({\bf r},t)+
2U_g[n_c({\bf r},t)+{n}_b({\bf
r},t)]\; ,
\label{veffbb}
\end{equation}
including the time-dependent driving
potential ${\cal U}({\bf r},t)$ and an axially symmetric confining potential
given for a harmonic trap by $V^{ext}_b({\bf r})=
m_{b}\omega_b^2(r_{\perp}^2+\epsilon_b^2 z^2)/2$. 
In Eqs. (\ref{veffab}) and (\ref{veffbb}),
$n_c({\bf r},t)=|\Phi({\bf r},t)|^2$ and 
$U_g=4\pi\hbar^2 a_{bb}/m_b$ is the boson-boson
interaction parameter in terms of the $s$-wave scattering 
length $a_{bb}$
and of the boson mass $m_b$.

Once the algorithm to solve Eqs. (\ref{eqc}) and (\ref{eqf}) is
implemented, it is easily extended to a
mixture of condensed bosons and a fermionic cloud  
in the collisionless regime. In this case, 
the effective mean-field potentials become
\begin{equation}
V^{eff}_{c,f}({\bf r},t)=V^{ext}_b({\bf r})+
{\cal U}({\bf r},t)+U_gn_c({\bf r},t)+
U_f n_f({\bf r},t)
\label{veffaf}
\end{equation}
and
\begin{equation}
V^{eff}_{f}({\bf r},t)=V^{ext}_f({\bf r})+{\cal U}_f({\bf r},t)+
U_fn_c({\bf r},t)\; .
\label{veffbf}
\end{equation}
Here, $V^{ext}_f({\bf r})=
m_{f}\omega_f^2(r_{\perp}^2+\epsilon_f^2 z^2)/2$ and 
${\cal U}_f({\bf r},t)$ are the external trapping and driving potentials acting
on the fermions and $U_f=2\pi\hbar^2 a_{bf}/m_r$ is the boson-fermion 
coupling constant
with $a_{bf}$ the boson-fermion $s$-wave scattering length and 
$m_r=m_bm_f/(m_b+m_f)$, $m_f$ being the fermion atomic mass.
Notice that fermion-fermion collisions in the $s$-wave channel
are effectively suppressed by the Pauli principle in a dilute gas
of spin-polarized fermions, as is relevant to current experiments
on boson-fermion mixtures.

\subsection{Validity of the model}
To summarize, Eqs. (\ref{eqc}) and (\ref{eqf}) 
describe the coupled dynamics of a Bose-Einstein condensate and
a bosonic or fermionic cold-atom cloud. In the former case 
the potentials $V^{eff}_{c,b}$ and $V^{eff}_{b}$
are used, and in the latter these are replaced by 
$V^{eff}_{c,f}$ and 
$V^{eff}_{f}$. 
The range of validity of this approach is in principle 
limited to: (i) slowly varying
space- and time-dependent external potentials, which allows a
low-order gradient expansion of the equations of 
motion for the one-body density matrix; 
(ii) a collisionless regime, which allows 
expansion of the self-energies to first order
in the strength of the atomic interactions; and (iii) 
not too low temperature, so that the anomalous averages can
be neglected.  

We implement below a numerical method to solve Eqs. (\ref{eqc}) 
and (\ref{eqf}) and test it against  
dynamical behaviors in a one-component fermionic 
cloud and in clouds of either fermions or thermal bosons 
accompanied by a condensate. Of course, the role of the statistics 
enters at this level only from the initial distribution of the
particles in phase space. We thus turn to discuss the
equations that we use to determine the initial
conditions
for the subsequent time evolution.

\subsection{Semiclassical description of the equilibrium state}

Before proceeding to present the algorithm for our
dynamical simulations we briefly recall 
the basic steps that we take in 
preparing the initial state of the gas in thermodynamic equilibrium
and in evaluating the corresponding densities of the condensate and of the
fermionic or bosonic cloud. We refer the reader to
Ref. \cite{MCT,Madda} for the details of the theory and for a 
discussion of the excellent agreement that it yelds with thermodynamic
data on Bose-Einstein condensed gases, under conditions of temperature
and dilution that will be verified in the calculations reported in Section
IV below. 

The equilibrium condensate density is calculated 
within the Thomas-Fermi 
approximation, which amounts to neglecting the kinetic energy term in the
GPE.
Its validity is ensured 
whenever the average mean-field energy $U_gn_c$ is much larger than 
the typical confining energy.
In the case of harmonic confinement the condition 
$N_c a_{bb}/a_{ho}\gg 1$ is required, where $N_c$ is the number of atoms in
the condensate,
$a_{ho}=(\hbar/m_b\overline{\omega}_b)^{1/2}$ is the harmonic
oscillator length and
$\overline{\omega}_b=\omega_b\epsilon^{1/3}$
is the geometric average of the trap frequencies.
The equilibrium density profile of the condensate is given by 
\begin{equation}
n_c({\bf r})=\frac{1}{U_g}\,\left[\mu_b-V^{ext}_{b}({\bf r})-k_{b,f}
 n_{b,f}({\bf r})\right]\,
\theta(\mu_b-V^{ext}_{b}({\bf r})-k_{b,f}
n_{b,f}({\bf r}))
\end{equation}
where $k_{b}=2U_g$, $k_{f}=U_f$ and
$\mu_b$ is the chemical potential for the bosons.

For the equilibrium cloud density of bosons or fermions
we adopt the semiclassical Hartree-Fock scheme~\cite{MCT}. This choice
is justified as long as the gas is
in a very dilute regime, 
a condition which is usually met in current experiments.
In this approximation we have
\begin{equation}
n_{b,f}({\bf r})=\int \frac{{\rm d}{\bf p}}{(2\pi)^3}
\left\{\exp{\left[
\frac{1}{k_BT}\left(\frac{p^2}{2m_{b,f}}+V^{eff}_{b,f}
({\bf r})-\mu_{b,f}\right)\right]}\mp1\right\}^{-1},
\label{thermal}
\end{equation}
with
$V^{eff}_{b}({\bf r})$ and $V^{eff}_{f}({\bf r})$ 
determined by the confining potentials supplemented
by static mean-field interaction term as
in Eqs. 
(\ref{veffbb}) or (\ref{veffbf}). 

The chemical potentials $\mu_b$ and $\mu_f$ are determined from 
the total numbers of bosons
and fermions. In the case of a bosonic thermal cloud, $\mu_b$
is fixed by the relation
\begin{equation}
N_b=\int {\rm d}{\bf r}
\left[n_c({\bf r})+n_b({\bf r})\right]\,.
\end{equation}
For a fermionic cloud the chemical 
potential is
determined from the total number of fermions,
\begin{equation}
N_f=\int {\rm d}{\bf r}\, N_f({\bf r})\,.
\label{closure}
\end{equation}
These equations complete the self-consistent closure of the model in the
initial equilibrium state.

\section{The numerical method}\label{Method}
In this section we present the numerical procedure that we have used
to solve the system of Vlasov-Landau and Gross-Pitaevskii equations.
Since most experimental setups are invariant
under rotation in the azimuthal plane, we use cylindrical coordinates
$\{r,z\}$. The wavefunction $\Phi$ is discretized on a two-dimensional 
grid of $N_r\times N_z$ points, which are uniformly distributed in a box of 
sizes $r_{max}\times2z_{max}$: that is, $\Phi_{jk} \equiv \Phi(r_j,z_k)$
with
\begin{equation}
\left\{
\begin{array}{lr}
r_j = (j-1) \Delta r& \;\;\;(j=1,\dots, N_r)\\
z_k = -z_{max}+(k-1) \Delta z&\;\;\;(k=1,\dots,N_z)\\
\end{array}\right.
\end{equation}
$\Delta r$ and $\Delta z$ being the steps in the two space variables.
The particle distributions are discretized by means of a set
of $P$ computational particles,
\begin{equation}
f({\bf p},{\bf r},t) \rightarrow f_{P} ({\bf p},{\bf r},t) \equiv
\sum_{i=1}^{P} \delta({\bf r}-{\bf r}_i(t)) \delta({\bf p}-{\bf p}_i(t))
\end{equation}
where the $6 P$ phase-space coordinates
${\bf r}_i(t)$ and ${\bf p}_i(t)$ represent the actual numerical unknowns
entering  the VLE.
They obey Newton's equations,
\begin{equation}
\left\{
\begin{array}{ll}
\dfrac{d {\bf r}_i}{dt} = {\bf p}_i/m\\
\dfrac{d {\bf p}_i}{dt} = {\bf F}_i
\end{array}\right.
\end{equation}
where ${\bf F}_i$ collects all forces acting on 
the $i$-th computational particle.
As a result, at each time step we have to solve for
$N_G=N_r N_z$ grid unknowns coupled to $6 P$ discrete
particle coordinates.
In order to keep the systematic signal reasonably above the noise level,
each grid cell contains of the order of $10-100$ computational particles.
As a result, $P \sim 10N_G-100 N_G$.

\subsection{Propagation of the Gross-Pitaevskii equation}

The GPE is advanced in time by an explicit finite-difference
method based on a non-staggered variant of the Visscher method 
\cite{visscher}.
The full details are given in original papers
\cite{PRE,PREGS} and  
here we shall just review the main ingredients of the algorithm.
The basic idea is to advance the real and imaginary
parts of the wavefunction, say $A$ and $B$, in alternating 
steps as follows: 
\begin{equation}
\left\{
\begin{array}{ll}
A_{jk}^{n+1}-A_{jk}^{n-1}=-2 [K_{jk}(B) + V_{jk}^{n}(B)] \Delta t\\
B_{jk}^{n+1}-B_{jk}^{n-1}= 2 [K_{jk}(A) + V_{jk}^{n}(A)] \Delta t
\end{array}\right.
\label{eq15}
\end{equation}
where $n=1, \dots, N_T$ labels the discrete time sequence $t_n=n \Delta t$.
The scheme is initiated as follows. The initial conditions specify
the values of $A$ and $B$ at $n=0$ and subsequently
a first-order Euler step provides their values at $n=1$. 
With these values available, all later steps $n=2,3, \dots, N_T$
are taken by using the above set of equations.

In Eq.~(\ref{eq15}) $K_{jk}$ is the kinetic energy operator,
\begin{equation}
K_{jk}(\Phi) =-\frac{\hbar^2}{2m}[ 
\frac{\Phi_{j-1,k}}{{\Delta r}^2}+
\frac{\Phi_{j+1,k}}{{\Delta r}^2}+
\frac{\Phi_{j,k-1}}{{\Delta z}^2}+
\frac{\Phi_{j,k+1}}{{\Delta z}^2}-
2(\frac{1}{{\Delta r}^2}+\frac{1}{{\Delta z}^2}) 
\; \Phi_{j,k}]
\end{equation}
and
\begin{equation}
V_{jk}(\Phi)=V^{eff}(r_j,z_k;n_{jk}) \Phi_{jk}
\end{equation}
is the potential energy operator, including both external and
self-consistent interaction terms.
The self-consistent potential requires the specification of the
particle density $n_{jk}$ at each node of the spatial grid.
This is obtained by convoluting the discrete 
particle distribution,
\begin{equation}
n_{jk}(t) = \sum_{i \in C_{jk}} W_{jk,i} f_i(t)
\end{equation}
where $C_{jk}$ is the grid cell centered in $r_{j+1/2}=(r_j+r_{j+1})/2$, and 
$z_{k+1/2}=(z_k+z_{k+1})/2$, while $f_i=1$ if the particle belongs to 
$C_{jk}$ and $f_i=0$ otherwise.
The factor $W_{jk,i}$ weights the contribution
of particle $i$ to the density at the grid-point $(r_j,z_k)$. 

\begin{figure}
\centerline{\subfigure[]{\epsfig{file=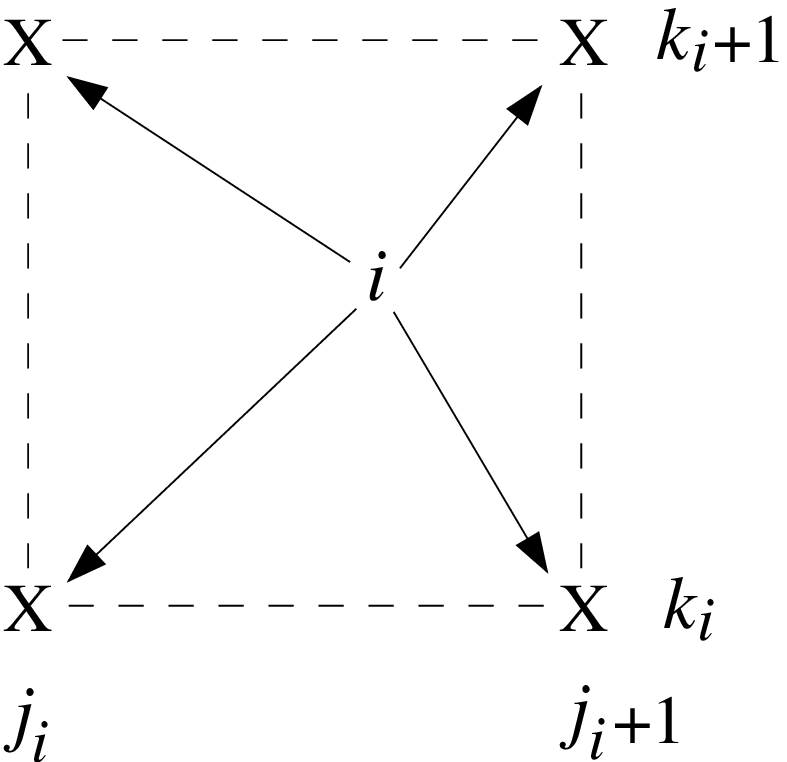,width=0.35\linewidth}}
\hspace{1.5cm}
\subfigure[]{\epsfig{file=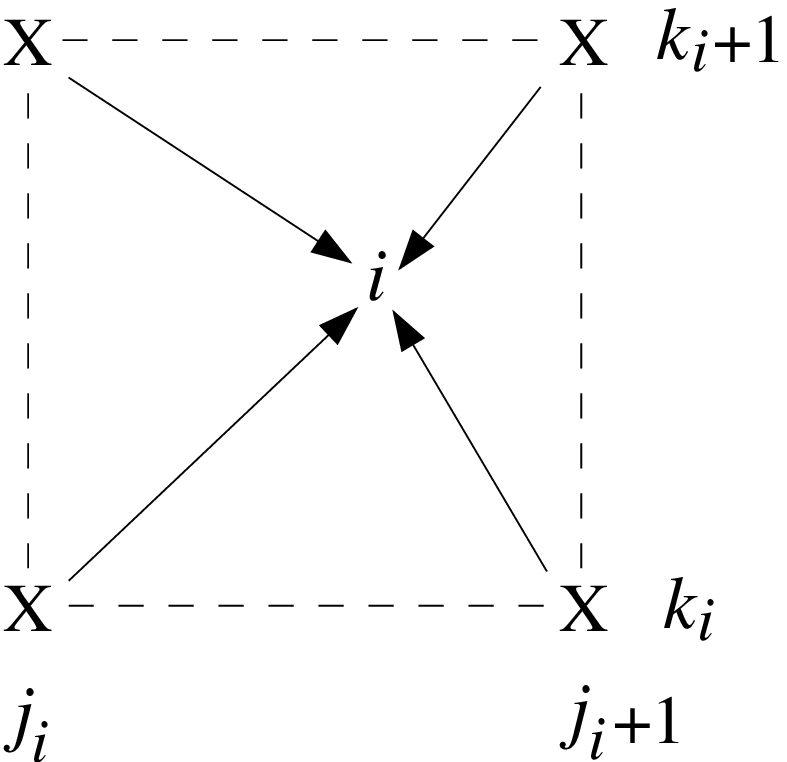,width=0.35\linewidth}}}
\caption{Bilinear CIC interpolation to weight the contribution of 
particle $i$ to the density at grid-point $(r_j,z_k)$ (a) and to weight
the effect of the forces at grid-point $(r_j,z_k)$ on particle
$i$ (b).}
\label{sauro1}
\end{figure}

We adopt a bilinear "cloud-in-cell" (CIC) interpolator 
(see fig.~\ref{sauro1}a), which yields
the scattering (particle-to-grid) rule \cite{HOCKNEY}
\begin{equation}
W_{jk,i} = e_i(r_j) e_i(z_k)\;\;\;
(j=j_i,j_i+1,\;\;\;k=k_i,k_i+1).
\label{eq19}
\end{equation}
Here $(j_i,k_i)$ identifies the lower-left corner 
of the grid cell to which the $i$-th particle belongs.
In Eq.~(\ref{eq19}) $e_i(r)$ and $e_i(z)$ are one-dimensional 
piecewise linear splines centered on the discrete
particle position $(r_i,z_i)$,
\begin{equation}
e_i(r) =\left\{
\begin{array}{r}
1+\dfrac{(r-r_i)}{\Delta r}\qquad\;\; \;(r_i-\Delta r<r<r_i)\\
1-\dfrac{(r-r_i)}{\Delta r}\qquad \;\;\;(r_i<r<r_i+\Delta r)
\end{array}\right.
\end{equation}
and similarly for $e_i(z)$.

\subsection{Propagating the Vlasov-Landau equation}

The VLE is advanced in time by a standard Particle-in-Cell
(PIC) method \cite{HOCKNEY}.
In a modified Verlet time-marching scheme, we obtain the
following set of discrete algebraic equations:

\begin{equation}
\left\{
\begin{array}{lll}
r_i(t+\Delta t)=r_i(t)+v_{ri}(t) \Delta t + a_{ri}(t) {\Delta t}^2/2\\
(r_i \theta_i)(t+\Delta t)=(r_i \theta_i)(t)+
v_{\theta i}(t) \Delta t + a_{\theta i}(t) {\Delta t}^2/2\\
z_i(t+\Delta t)=z_i(t)+v_{zi}(t) \Delta t + a_{zi}(t) {\Delta t}^2/2\\
v_{ri}(t+\Delta t)=v_{ri}(t)+ a_{ri}(t) {\Delta t}\\
v_{\theta i}(t+\Delta t)=v_{\theta i}(t)+ a_{\theta i}(t) {\Delta t}\\
v_{zi}(t+\Delta t)=v_{zi}(t)+ a_{zi}(t) {\Delta t}
\end{array}\right.
\end{equation}
Here $a_r,a_{\theta},a_z$ and $v_r,v_{\theta},v_z$
are the three components of the acceleration and of the velocity
along the $(r,\theta,z)$ coordinates.
The advantage of the modified Verlet time marching is that 
fourth-order accuracy is preserved while synchronously
keeping the coordinate and
momentum degrees of freedom on the same sequence of discrete
times. 

The algorithm is standard except
for the specification of the self-consistent coupling, namely the
force due to the density gradients.
In the first place, we form density gradients from auxiliary values 
of the density field $n_{j+1/2,k+1/2}$ at cell centers,
\begin{equation}
\left\{
\begin{array}{ll}
Gr_{ij} = [n_{j+1/2,k-1/2}-n_{j-1/2,k-1/2}+
n_{j+1/2,k+1/2}-n_{j-1/2,k+1/2}]/(2 \Delta r)\\
Gz_{ij} = [n_{j-1/2,k+1/2}-n_{j-1/2,k-1/2}+
n_{j+1/2,k+1/2}-n_{j-1/2,k-1/2}]/(2 \Delta z)
\end{array}\right.
\end{equation}
The azimuthal acceleration is zero 
in cylindrical symmmetry.
Next, the grid-forces are evaluated on the discrete particle locations,
this being the inverse of the scattering operation discussed in 
the previous section
(see fig.~\ref{sauro1}b).
The grid-to-particle convolution is
\begin{equation}
{\bf F}_{i} = \sum_{s=0}^1 W_{i,(j+s,k+s)} {\bf F}_{j+s,k+s}. 
\end{equation}
In order to avoid spurious self-forces we use
again CIC-interpolation, which amounts to using
the same weighting function as for the GPE in the grid-to-particle 
scattering rule,
\begin{equation}
W_{i,jk} = e_{j}(r_i) e_{k}(z_i).
\end{equation}
With the force/acceleration field transferred to the particle locations,
everything is set to march the VLE in time.

\subsection{Boundary conditions}

The conditions imposed on the wavefunction are (i) periodicity along the 
$z$ coordinate, (ii) symmetry (zero radial gradient) at $r=0$, and
(iii) vanishing at the outer radial boundary $r=R$.
For the discrete particle distribution we have again periodicity
along $z$ and specular reflection at the outer radial boundary.
Specular reflection means that a particle flying from, say, $r_{\alpha}<R$ to
$r_{\beta}>R$ is replaced by a particle at $r=2R-r_{\beta}$ 
with inverted radial speed.
Since the particle trajectories are tracked in a three-dimensional
cylindrical coordinate frame, the $r=0$ axis requires
no special treatment.

\subsubsection{Time-step considerations}

The GPE and the VLE are advanced on the same discrete
time sequence. This maximizes simplicity, but implies that the time-step
is controlled by the fastest process at work, which usually is 
the self-consistent potential acting upon the condensate wavefunction.
Better efficiency can be achieved by {\it sub-cycling} the time-stepper,
namely by advancing the slowest equation (say the VLE) only every
$\Delta t_{VL}/\Delta t_{GP}$ steps, $\Delta t_{VL}$ and $\Delta t_{GP}$ 
being the largest time-steps allowed by the stability conditions 
on the two equations.
The maximum time-step for the GPE solver is estimated from
\begin{equation}
\Delta t_{GP} (C_1 \frac{\hbar}{m \delta^2} + C_2 \hbar\dfrac{V_M}{\hbar}) < 1
\end{equation}
where $\delta$ is a typical mesh size, $V_M$ is the maximum value
of the potential and $C_1$ and $C_2$ are two coefficients $O(1)$ which
depend on geometry and dimensionality.
We note here the concurrent effects of quantum diffusion (kinetic energy)
and scattering/absorption (potential energy).

The maximum time-step for the VLE solver is estimated from
\begin{equation}
\Delta t_{VL} \frac{v_{max}}{\delta} < 1
\end{equation}
where $v_{max}$ is the maximum speed in the velocity grid, of the order
of the Fermi velocity for fermions and of the thermal velocity 
for bosons.
Under ordinary conditions the  kinetic energy contributions dominate 
over those from the potential energy, 
so that the condition for the GPE to be the time-limiting
section of the code takes the form of a numerical ``uncertainty principle'',
$m v_{max} \delta > \hbar$.
For the cases discussed in this work, this inequality
is generally fulfilled within a factor ten, so that sub-cycling 
is not compulsory. The inclusion of collisional interactions 
would make it mandatory.

\subsubsection{Radial singularity}

A source of potential trouble is the singularity in $r=0$,
which is known to affect all calculations in polar coordinates.
To date, singular factors $1/r$ are regularized by a simple numerical
cutoff $1/r \rightarrow 1/(r+r_c)$ with $r_c \sim 0.001 \Delta r$,
with a check that the physical results are virtually insensitive to the
specific value of $r_c$. 

Another undesirable side-effect of the cylindrical
geometry is the relative depletion of near-axis cells, which tend to
host fewer particles just because of the $r \Delta r$ volumetric effect.
On the other hand, this volumetric depletion is often more than compensated
by the physical behavior of the radial density, which is generally
largest at $r=0$. At any rate, the volumetric effect
can be readily disposed of by moving to a non-uniform grid
along the radial coordinate.

\subsubsection{Statistical noise}

Considerations of statistical accuracy
require of the order of a few tens of particles
per grid-point (or equivalently per grid-cell) 
in order to keep the noise-to-signal ratio 
below an acceptable threshold.
A practical consequence of this statistical accuracy
requirement is that the VLE part of the
computational scheme should be designed in such a way
as to evolve these tens of computational particles
in approximately the same amount of CPU time that it takes
the GPE solver to advance a single grid-point.

Another interesting consequence is that -
at variance from ordinary situations in (classical) rarefied-gas 
dynamics - the number of computational particles
in the simulation of Bose-Einstein condensates far
exceeds the number of physical atoms, typically
by a factor $10^3$ in our case.
As a result, each single computer simulation performs
{\it de facto} a built-in ensemble average over a 
set of about a thousand realizations.

\subsection{Procedure: preparing the initial state}
\label{Procedure}
The initial condition for the populations
of bosons or fermions in the cold-atom cloud is prepared as follows.
Particles are sampled from the probability
distribution functions (pdf)
\begin{equation}
\label{PDF}
f_{b,f} = \frac{e^{-\eta}}{1 \mp e^{-\eta}}
\end{equation}
where $\eta \equiv \beta (\mu_{b,f}-V_{eff}+p^2/2m)$
with $\beta = 1/k_B T$ the inverse temperature.

The initializiation procedure starts by assigning   
to each spatial cell centered about position ${\bf r}$
a corresponding amount of computational particles,
\begin{equation}
\Delta N({\bf r}) = n_{b,f}({\bf r}) \Delta V({\bf r})
\end{equation}
where $\Delta V ({\bf r})= 2 \pi r \Delta r \Delta z$ is the cell volume.
These particles must be sampled in momentum space
according to the pdf (\ref{PDF}).
Owing to the non-separability of the pdf, straightforward sampling
based on exact inversion is ruled out and more general - but
less efficient - accept/reject methods must be resorted to.

The particle momenta are sampled from the distribution function
using the standard Box-Mueller algorithm in three
dimensions for cylindrical coordinates\cite{PANG},
\begin{equation}
\left\{
\begin{array}{l}
p_{\perp}        = p_{max} r_1 \\
\xi    = 2 \pi r_2\\
p_z=p_{max}(2r_3-1)\\
\end{array}\right.
\end{equation}
where $r_1$, $r_2$ and $r_3$ are sampled from a uniform distribution
in the range $[0,1]$. The maximum momentum $p_{max}$ is taken to be
of order $2p_F$ for cold fermions and of order
$3m_b v_{th}$ for bosons, with $p_F=\hbar k_f$ being the Fermi
momentum and $v_{th}$ the average thermal velocity.
The particle momentum coordinates in the azimuthal plane are evaluated as
\begin{equation}
\left\{
\begin{array}{lll}
p_{r} &     = &p_{\perp} sin \xi\\
p_{\theta}& = &p_{\perp} cos \xi.\\
\end{array}\right.
\end{equation}

Finally, in order to avoid poor acceptance rates, 
the standard accept/reject test is performed
by comparing the pdf with the maximum value $f_{max}({\bf r})$ that it
can take in the cell at position
${\bf r}$, that is
\begin{eqnarray}
I\!f \; f(p_r,p_{\theta},p_z;{\bf r}) > r_4f_{max}({\bf r}): 
\; accept\nonumber\\
Else: \; reject
\end{eqnarray}
where $r_4$ is a random number uniformly distributed in
$[0,1]$.

\section{Physical applications}\label{Results}
\subsection{Expansion of cold fermions}
As a first test of the dynamical algorithm, we consider the expansion
of a cloud of cold fermions after release of the harmonic trap. The neglect of
the collisional integrals is a reliable approximation in this
context, since as already remarked the $s$-wave scattering between 
spin-polarized fermions is
suppressed by the Pauli principle.  

This problem  has
been analytically solved in Ref.~\cite{Renzo} under the assumption of
ballistic expansion. The time-evolution of the mean square radii is
found to be  
\begin{equation}
<r_{\perp}^2(t)>=\frac{1}{3N_f}E_{rel}{4\over m_f\omega_f^2}(1+\omega_f^2 t^2)
\label{fexpar}
\end{equation}
and
\begin{equation}
<z^2(t)>=\frac{1}{3N_f}E_{rel}{2\over
m_f(\epsilon_f\omega_f)^2}(1+(\epsilon_f\omega_f)^2 t^2)\; .
\label{fexpaz}
\end{equation}
In Eqs.~(\ref{fexpar}) and (\ref{fexpaz}) $E_{rel}$ is the so-called
release energy, namely the energy of the system with $N_f$ 
fermions after switching off
the trap, which amounts to one half of the total average energy. 
For a non-interacting fermion gas at temperature $T>0.2 T_F$ 
this is best  approximated by the classical relation 
$E_{rel}\simeq 3N_fk_BT/2$, while 
at lower temperature  it is given by $E_{rel}\simeq (3/4)N_f
E_F[1+(2\pi/3)(T/T_F)^2]$, with $E_F=k_B T_F=(6
N_f)^{1/3}\hbar\omega_f$ being the Fermi energy. 

We prepare the initial thermodynamic state (\ref{thermal}) for 
$N_f=1000$ $^{40}K$
atoms in an isotropic trap with $\omega_f=2\pi\times 15.92$ rad/s at
$T=0.55\; T_F\approx 7.6$ nK.
The chemical potential of the gas is  
$\mu_f=0.063\; E_F=1.14\;\hbar\omega_f$. 
 We use a $N_r\times N_z$ mesh of $201$x$401$ points
 with $1.6\cdot 10^6$ 
representative particles 
in a box of sizes $40$ and $80$ in units of
$a_{ho}=\sqrt{\hbar/(m_f\omega_f)}$ along the radial and axial directions,
respectively. The time-step in the dynamical simulation is 
$\omega_f\Delta t=10^{-4}$.

We then evaluate from the simulation runs the radial width 
\begin{equation}
\sigma_{r_\perp}(t)
=\sqrt{\int dx\,dy\, dz\, [(x(t)-<\!x(t)\!>)^2+(y(t)-<\!y(t)\!>)^2]\,
n_f({\bf r};t)}
\end{equation}
and the axial width 
\begin{equation}
\sigma_z(t)=\sqrt{\int dx\, dy\, dz\, (z(t)-<\!z(t)\!>)^2n_f({\bf r};t)}
\end{equation}
of the cloud as functions of time, after averaging over 
the density distribution $n_f({\bf r};t)$. Here, 
the center-of-mass coordinates are defined as 
$<\!x(t)\!>
=\int dx\, dy\, dz\, x \,n_f({\bf r};t)$ 
and similarly for $<\!y(t)\!>$ and $<\!z(t)\!>$. 
Of course, during free expansion the center-of-mass coordinates must 
remain unchanged: this property is used as a test of the numerical
method.

\begin{figure}
\centerline{\epsfig{file=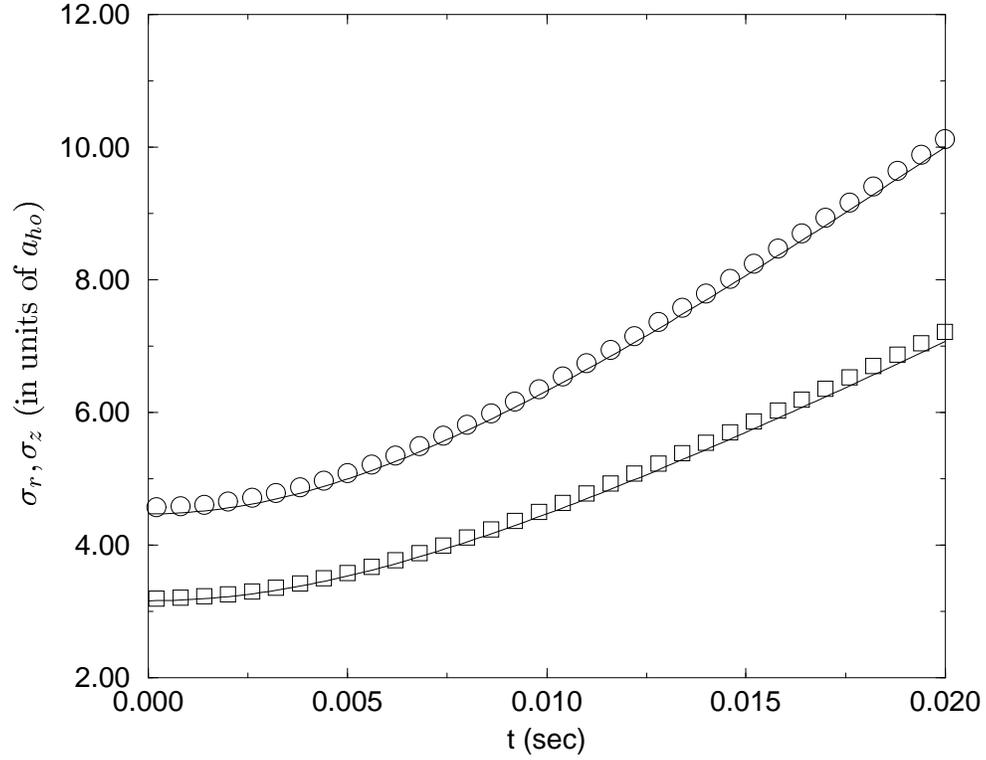,width=0.8\linewidth}}
\caption{Expansion of a cold fermionic cloud after release from a
harmonic trap: radial and axial
widths of the density
distribution (in units of $a_{ho}$)  as functions of time.  
Circles: $\sigma_{r_\perp}$; squares: $\sigma_z$; solid lines:
analytical expressions (\ref{fexpar}) and 
(\ref{fexpaz}) from Ref. \protect\cite{Renzo}.}
\label{renzofig1}
\end{figure}

\begin{figure}
\centerline{\epsfig{file=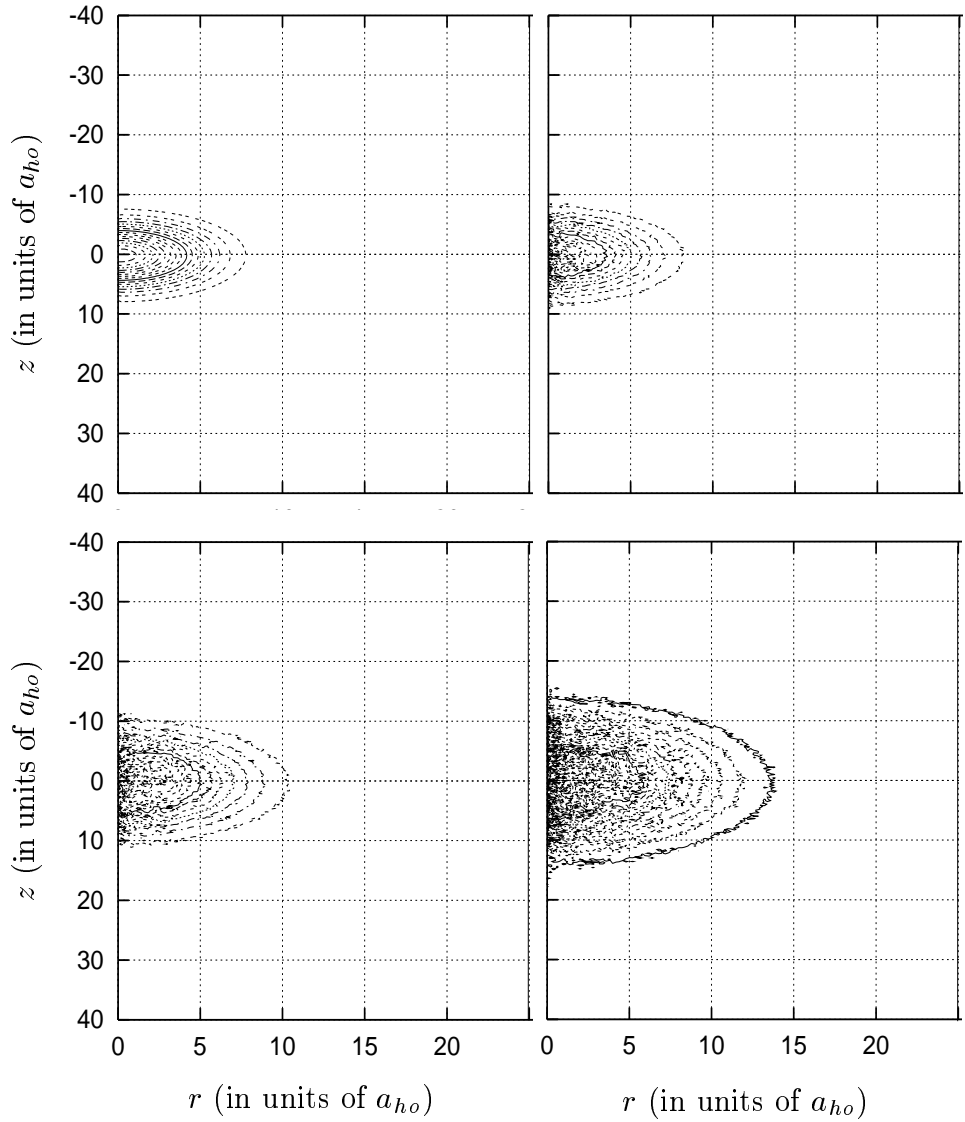,width=0.8\linewidth}}
\caption{Expansion of a cold fermionic cloud after release from a
harmonic trap: 
snapshots of the
density distribution, shown as contour plots. From top left to bottom right:
$t=0$, $5$, $10$ and $15$ ms. The axial and radial coordinates are 
in units of $a_{ho}$. }
\label{renzofig2}
\end{figure}

Fig.~\ref{renzofig1} shows that the calculated 
$\sigma_{r_\perp}$ (circles) and $\sigma_z$ (squares) 
agree with the results from the  
analytical expressions (\ref{fexpar}) and 
(\ref{fexpaz}) (solid lines), where the classical expression has been used 
for $E_{rel}$.
Snapshots of the density
profiles at selected times are shown
as contour plots in fig.~\ref{renzofig2}. 
The definition of the 
profile degrades in time because the number of particles
per cell drops during the expansion.

After this test of the reliability of the simulational method,
we proceed to use it in some original applications.

\subsection{Expansion of a mixture of a condensate and a cold-fermion cloud}

As a first novel application we look at the case in which a core of 
Bose-condensed atoms is present
inside
the dilute Fermi gas. We prepare a state with $N_f=1000$ $^{40}K$ atoms
and $N_c=10^5$ $^{39}K$ atoms in identical harmonic traps 
and at the
same temperature as for
the Fermi gas studied in Sec. IV.A. 
The scattering length which describes 
the interactions between the atoms in the
condensate is $a_{bb}=80\; a_{Bohr}$, while the
interspecies scattering length is $a_{bf}=40\; a_{Bohr}$.

The inclusion of the GPE algorithm at fixed mesh size normally 
requires shorter time-steps to mantain stability.
We make the choice of a thinner $501$x$1001$ 
mesh than in Sec. IV.A 
in order to keep the time-step at $\omega \Delta t=10^{-4}$, 
all other simulation
parameters remaining the same.

\begin{figure}
\centerline{\epsfig{file=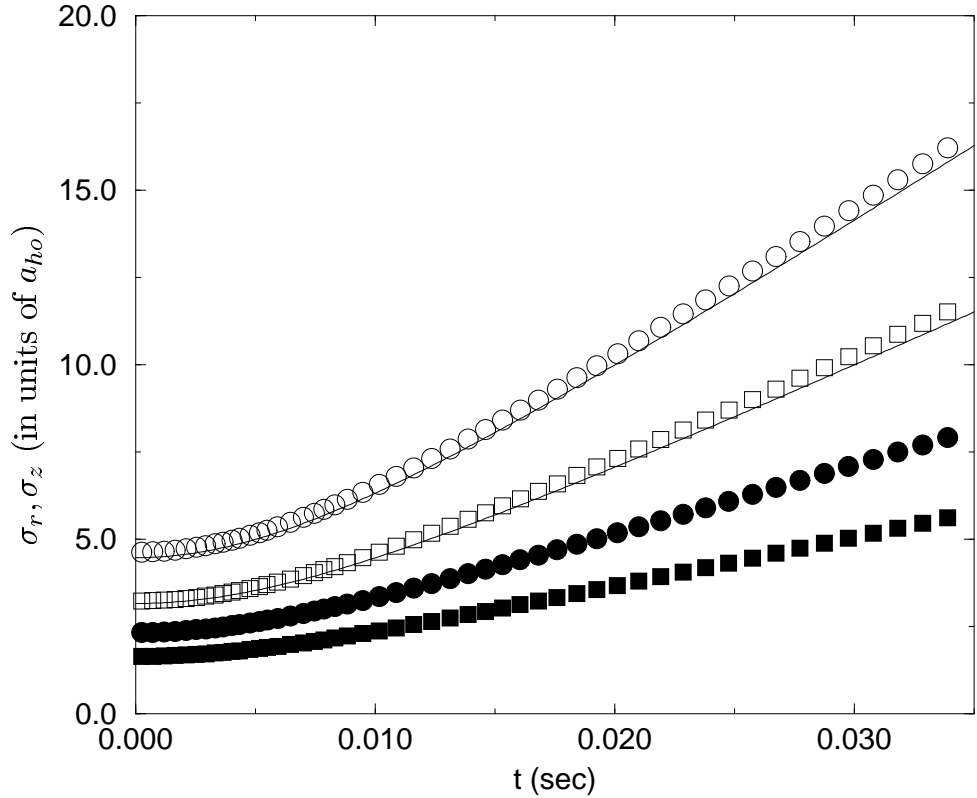,width=0.8\linewidth}}
\caption{Expansion of a cold fermionic cloud and an inner condensate core
after release from a harmonic trap: 
radial and axial widths of the density
distributions (in units of $a_{ho}$) as functions of time.  
Circles: $\sigma_{r_\perp}$; squares: $\sigma_z$; open symbols: fermionic 
cloud; filled symbols: condensate; solid lines:
analytical expressions (\ref{fexpar}) and 
(\ref{fexpaz}) from Ref.\protect\cite{Renzo}.}
\label{bfrenzofig1}
\end{figure}

The initial  state is characterized by  
$\mu_b=0.52 E_F=9.51\; \hbar\omega_f $ 
and $\mu_f =0.10E_F=1.83\; \hbar\omega_f $.  
We display in fig.~\ref{bfrenzofig1} the average widths 
$\sigma_{r_{\perp}}$ (circles) and $\sigma_z$ (squares) for both the fermionic
cloud (open symbols) and the condensate (filled symbols). The solid 
lines are the analytical solution for the ideal Fermi cloud as in
fig.~\ref{renzofig1}. It is seen that with the scattering lengths
of the $^{39}$K-$^{40}$K mixture the mean-field 
force of the inner condensate core on the outer fermionic cloud 
is not strong enough to sizeably affect the expansion of the latter.

\begin{figure}
\centerline{\epsfig{file=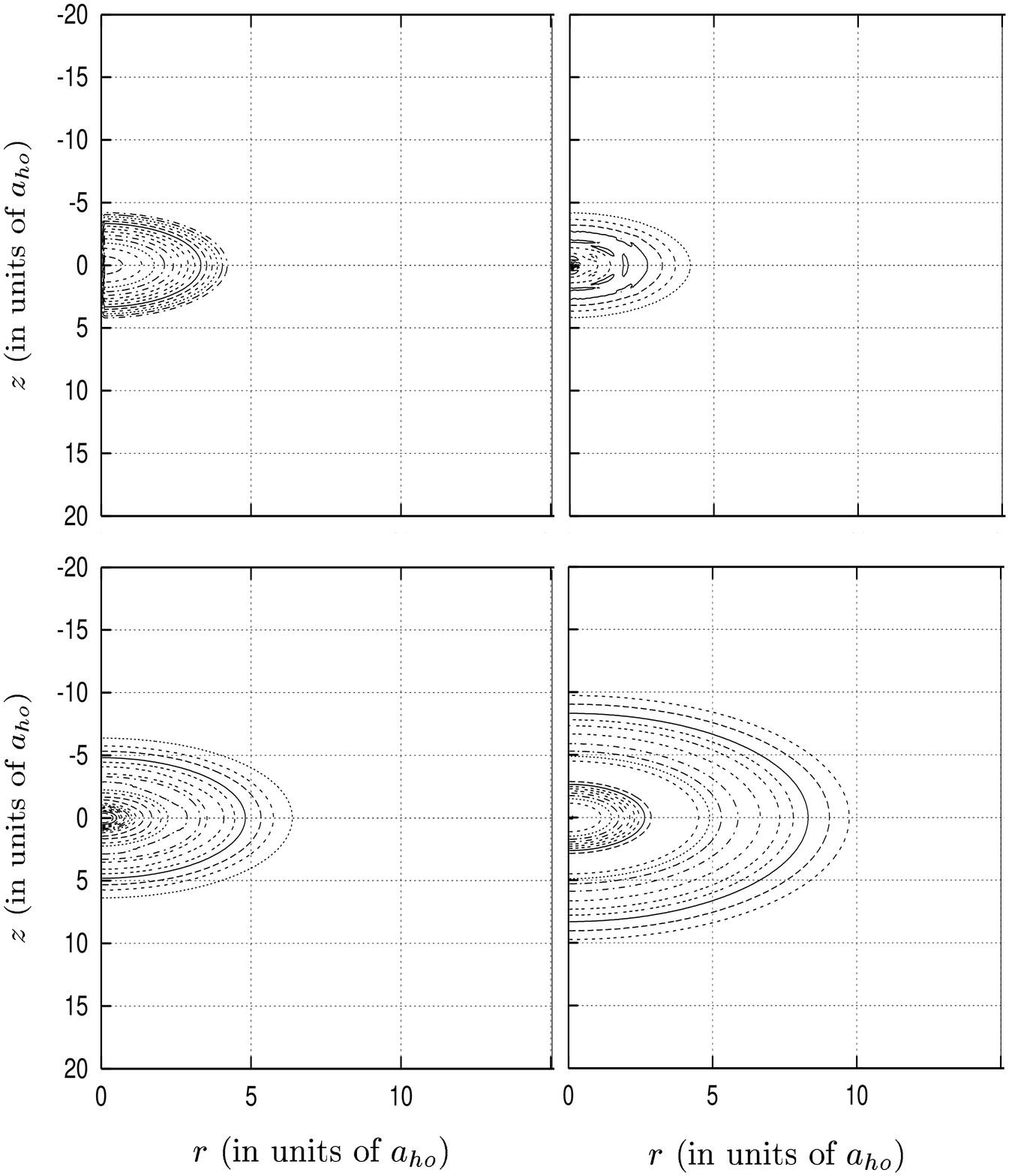,width=0.8\linewidth}}
\caption{Expansion of a cold fermionic cloud with an inner condensate core
after release from a harmonic trap: 
snapshots of the
condensate density distribution, shown as contour plots. From top left to bottom right:
$t=0$, $8.5$, $17$ and $25.5$ ms. The axial and radial coordinates are 
in units of $a_{ho}$. }
\label{bfrenzofig2}
\end{figure}

\begin{figure}
\centerline{\epsfig{file=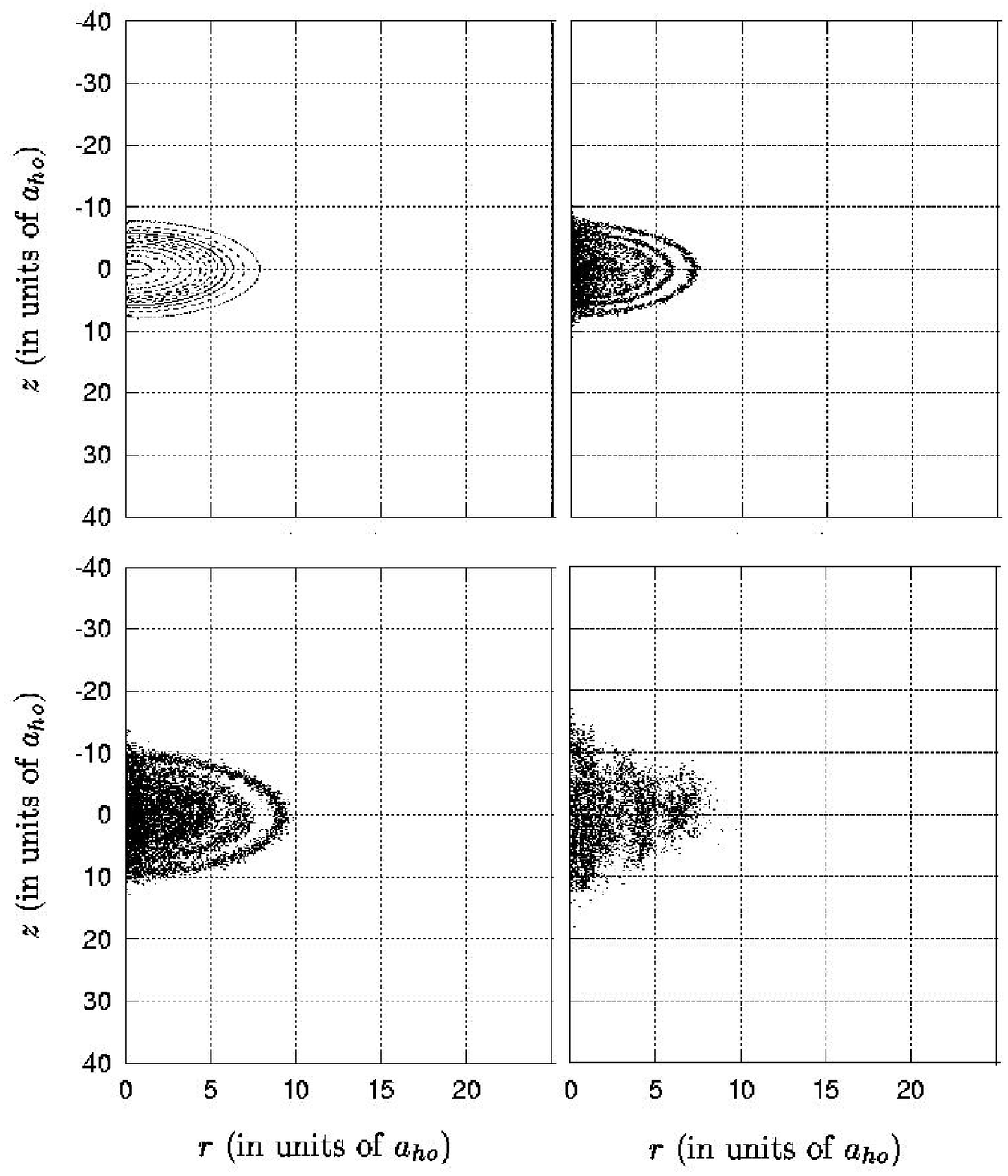,width=0.8\linewidth}}
\caption{Expansion of a cold fermionic cloud with an inner condensate core
after release from a harmonic trap: 
snapshots of the
fermionic density distribution, shown as contour plots. From top left to bottom right:
$t=0$, $8.5$, $17$ and $25.5$ ms. The axial and radial coordinates are 
in units of $a_{ho}$.}
\label{bfrenzofig3}
\end{figure}

Snapshots of the condensate density profiles
and of the fermionic cloud
at times $t=0$, $8.5$, $17$ and $25.5$
ms are displayed as contour plots 
in figs.~\ref{bfrenzofig2} and~\ref{bfrenzofig3}.
Comparison with those in
fig.~\ref{renzofig2}
shows that the reduced number of computational particles per cell tends to
increase the statistical noise. This degradation worsens
as the simulational
time elapses, 
as is evidenced by the last snapshot in fig.~\ref{bfrenzofig3}.

\subsection{Oscillations of Bose gases inside an optical lattice}
Here and in the following subsection we apply our
numerical method to study the dynamics of a Bose-Einstein condensate and a
thermal cloud of $^{87}$Rb
atoms at finite temperature inside a one-dimensional optical lattice.
The initial state is prepared by adding to the harmonic trap, described by
$V_{trap}({\bf r})=(1/2)m \omega^2(r_\perp^2+\epsilon^2 z^2)$,
a periodic potential given by $V_{latt}(z)=
\alpha E_R\sin^2(k_Lz)$, where 
$E_R=\hbar^2k_L^2/(2m)$ is the recoil energy
and $k_L=2\pi/\lambda$ is the wave number 
of the laser beam which creates an optical lattice
with period $\pi/k_L$ in the axial direction. 

Such a system, which has been
realized at LENS \cite{LENS} and examined numerically at $T=0$ 
by two of us~\cite{LENS,EPL},
shows a rich variety
of dynamical behaviors. 
Thus, the study of the sloshing-mode oscillations of an almost pure
condensate with $N=3\cdot 10^5$ atoms in a lattice with 
$\alpha=1.6$
shows that superfluidity is superseded by dissipation as the initial displacement
of the condensate away from the harmonic-trap center is increased.
This behavior is quantitatively understood as a gradual  
destruction of superfluidity {\it via} emission of sound waves
in the periodically modulated inhomogeneous medium\cite{LENS}. 
Below the dissipative threshold, on the other hand, the oscillatory
motion of the condensate through the optical lattice can be mapped into the dynamics
of superconducting carriers through a weak-link Josephson junction~\cite{EPL}.
This implies the possibility of observable resonances 
and of multimode behavior.

Here we extend the above numerical studies to contrast the oscillations
of a condensate with the motions of a thermal cloud.
We prepare initial states for the two cases $\alpha=1$ and $\alpha=5$,
both at
$T=0$ for the BEC~\cite{nonlin} and for the thermal cloud at temperature
$T$ above the critical temperature $T_c$.
We give an initial displacement $\Delta z=42.6\;\mu$m to the trap center
and follow the subsequent dynamics with a time-step $\omega\Delta t$
of order $10^{-5}$.

\begin{figure}
\centerline{\epsfig{file=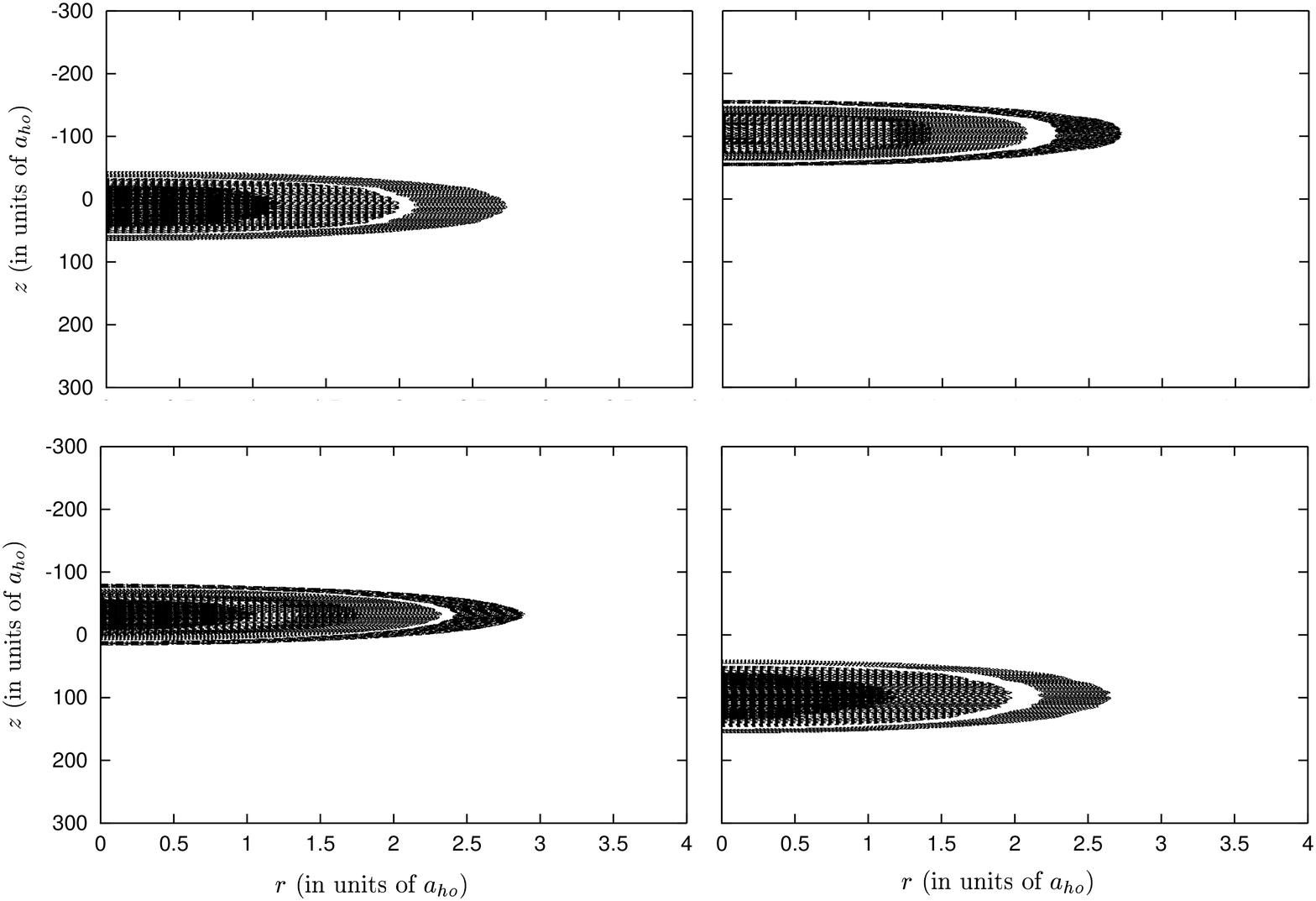,width=0.95\linewidth}}
\caption{Oscillations of a condensate in a
harmonic plus shallow optical-lattice potential: snapshots of the 
density distribution, shown as contour plots
From top left to bottom right:
$t=23.3$, $46.6$, $69.9$ and $93.2$ ms. 
The axial and radial coordinates are 
in units of $a_{ho}$.}
\label{osc_1_cond}
\end{figure}

\begin{figure}
\centerline{\epsfig{file=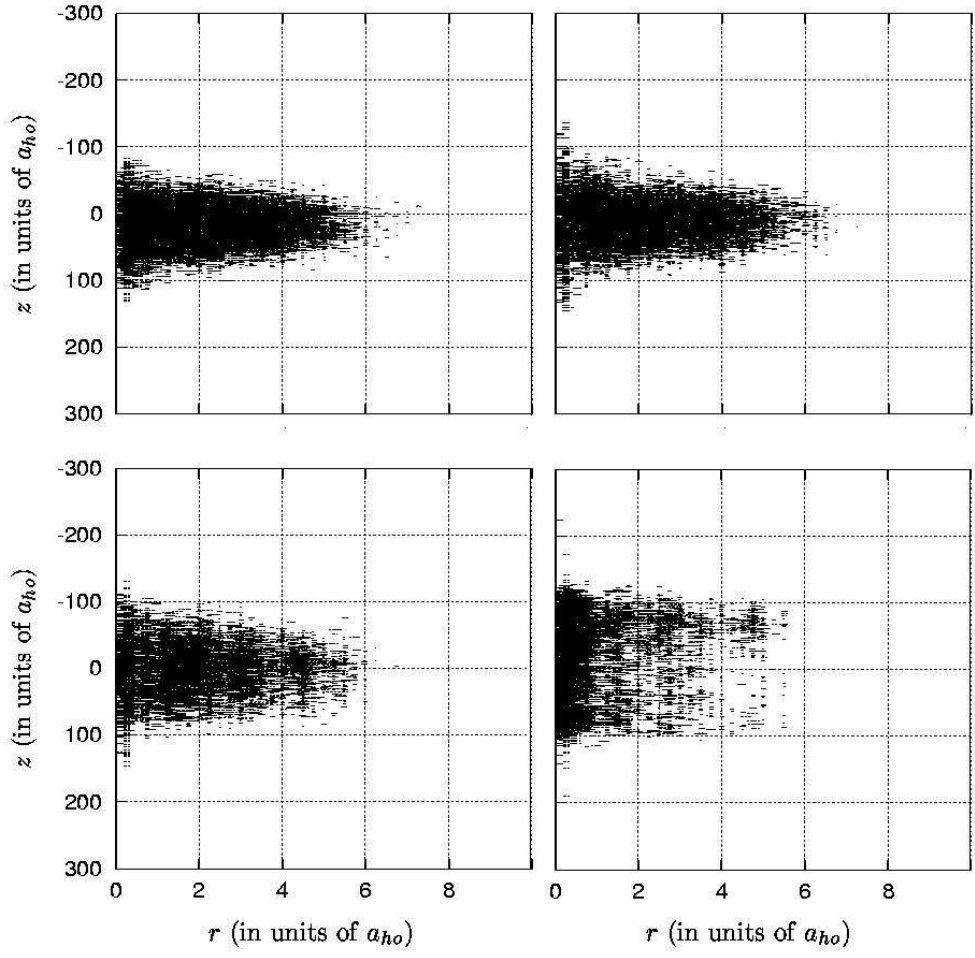,width=0.8\linewidth}}
\caption{Oscillations of a bosonic thermal cloud in a
harmonic plus shallow optical-lattice potential: snapshots of the 
density distributions, shown as contour plots.
From top left to bottom right:
$t=6$, $12$, $18$ and $24$ ms. 
The axial and radial coordinates are 
in units of $a_{ho}$.}
\label{osc_1_term}
\end{figure}

\begin{figure}
\centerline{\epsfig{file=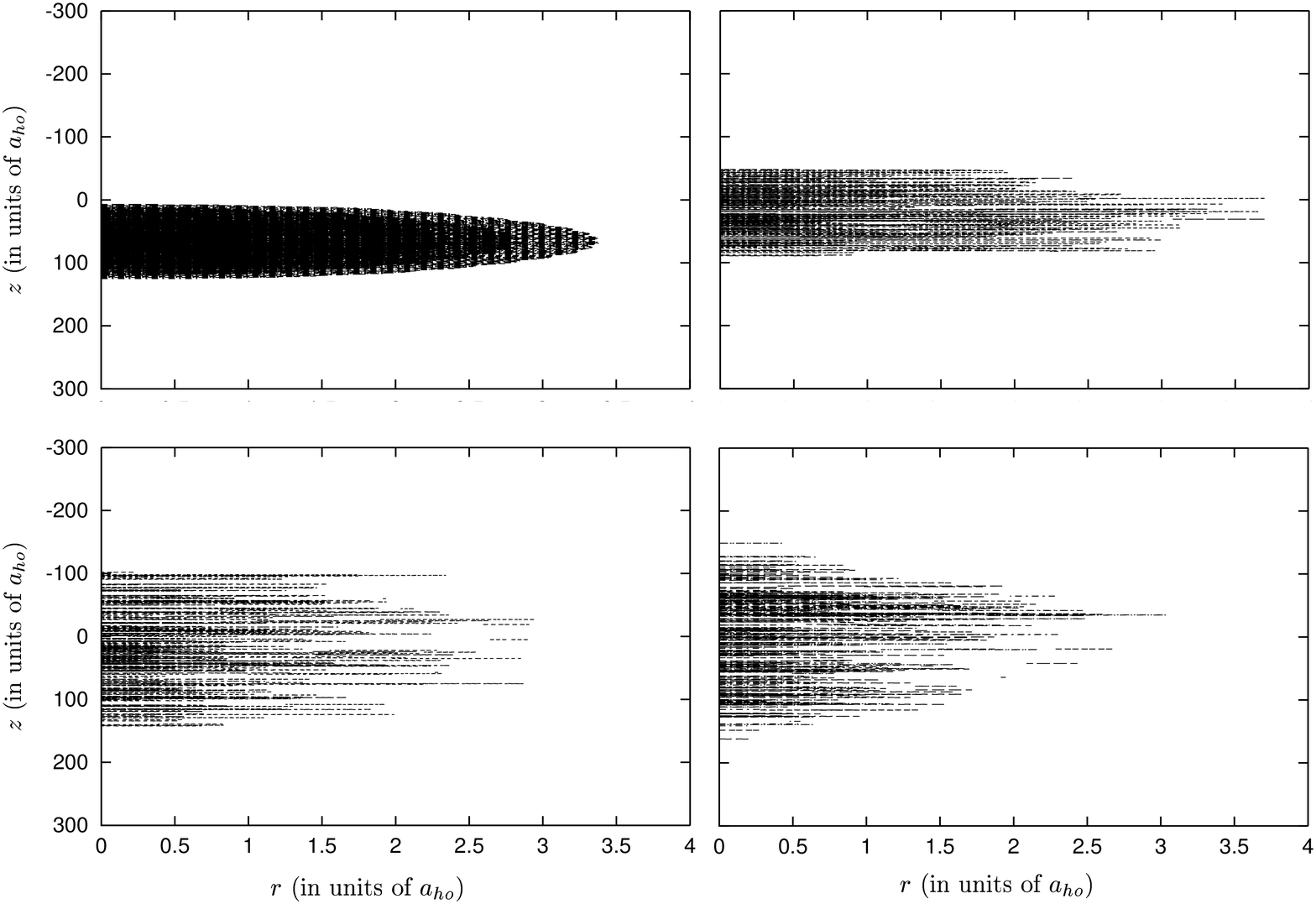,width=0.95\linewidth}}
\caption{Oscillations of a condensate in a
harmonic plus deep optical-lattice potential: snapshots of the 
density distributions, shown as contour plots.
From top left to bottom right:
$t=23.3$, $46.6$, $69.9$ and $93.2$ ms.
The axial and radial coordinates are 
in units of $a_{ho}$.}
\label{osc_5_cond}
\end{figure}

\begin{figure}
\centerline{\epsfig{file=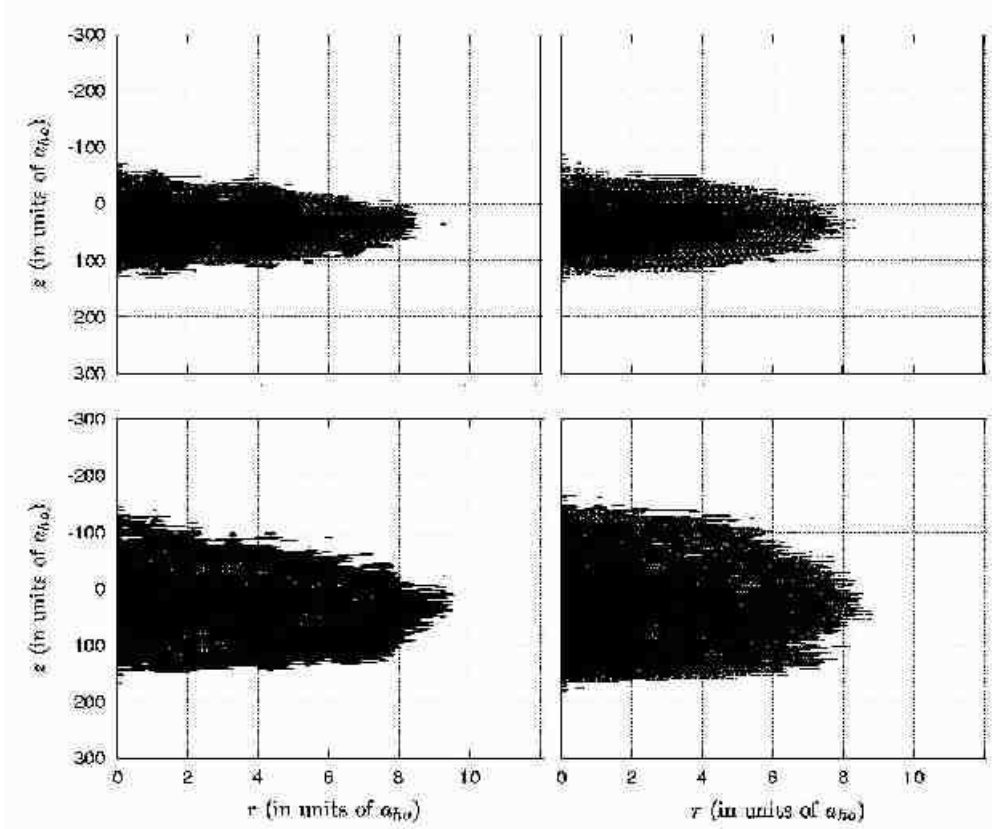,width=0.8\linewidth}}
\caption{Oscillations of a bosonic thermal cloud in a
harmonic plus deep optical-lattice potential: snapshots of the 
density distributions, shown as contour plots.
From top left to bottom right:
$t=2.8$, $5.6$, $8.4$ and $11.2$ ms. 
The axial and radial coordinates are 
in units of $a_{ho}$.}
\label{osc_5_term}
\end{figure}

The snapshots of the atomic density show that
for $\alpha=1$ (see Figs.~\ref{osc_1_cond} 
and ~\ref{osc_1_term}) the condensate behaves as a superfluid
executing harmonic oscillations at a frequency 
equal to the trap frequency, while the thermal cloud 
at $T>T_c$ diffuses away in a quarter of a period.
For $\alpha=5$ (see Figs.~\ref{osc_5_cond} and~\ref{osc_5_term})
the condensate breaks instead into fragments as it   
attempts to perform the first oscillation, and after a period 
its center of mass becomes localized at 
the bottom of the harmonic well. In the same setup the 
thermal cloud becomes localized at the center of the trap
in one tenth of a period and spreads out.

\begin{figure}
\centerline{\epsfig{file=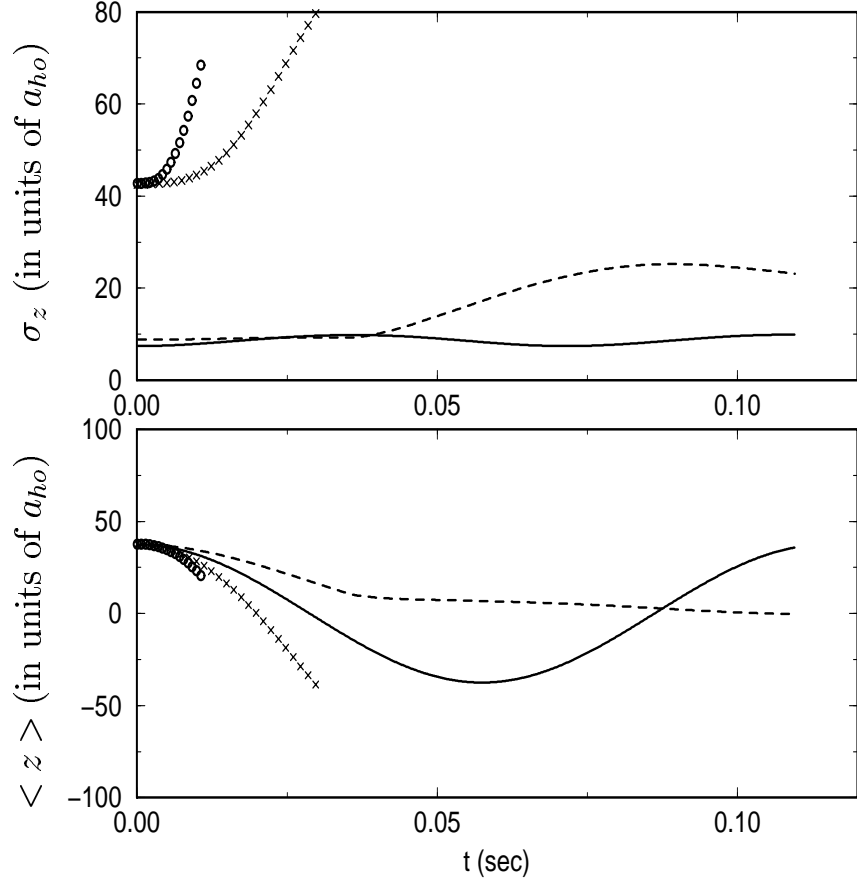,width=0.7\linewidth}}
\caption{Oscillations of a condensate and a bosonic thermal cloud in a
harmonic trap plus optical-lattice potential with: 
axial center-of-mass coordinate and
average axial width of the density
distributions (in units of $a_{ho}$) as functions of time. 
Continuous line, condensate with $\alpha=1$; dashed line, condensate with 
$\alpha=5$; crosses for the thermal cloud with $\alpha=1$ and 
circles for the thermal cloud with $\alpha=5$.}
\label{zmed}
\end{figure}

Figure~\ref{zmed} gives a clear picture of these behaviors by reporting the axial 
center position and width of the condensate and of the thermal cloud
as functions of time in the two cases.

\subsection{Expansion of a Bose-condensed gas in an optical lattice}
In our last study we look at the expansion of a Bose-Einstein condensate and 
its thermal cloud, which initially reside in a harmonic well and a superposed
optical-lattice potential. The external potentials are characterized by
parameters typical of an experiment at LENS~\cite{pedri}, namely
$\omega=2\pi\cdot 90$ rad/s, $\epsilon=8.9/90$, 
$2\pi/k_{L}=795$ nm and $\alpha=5$.
The condensate contains $6935$ $^{87}$Rb atoms and the
thermal cloud is composed of 
$3065$ atoms: the 
temperature of the gas is $T=86\; nK=0.24\; E_R/k_B$ and
its chemical potential is $\mu=5.86\;\hbar\omega =0.14\; E_R$.
We use a mesh of 111$\times$2801 points with 308000 representative
particles.

\begin{figure}
\centerline{\epsfig{file=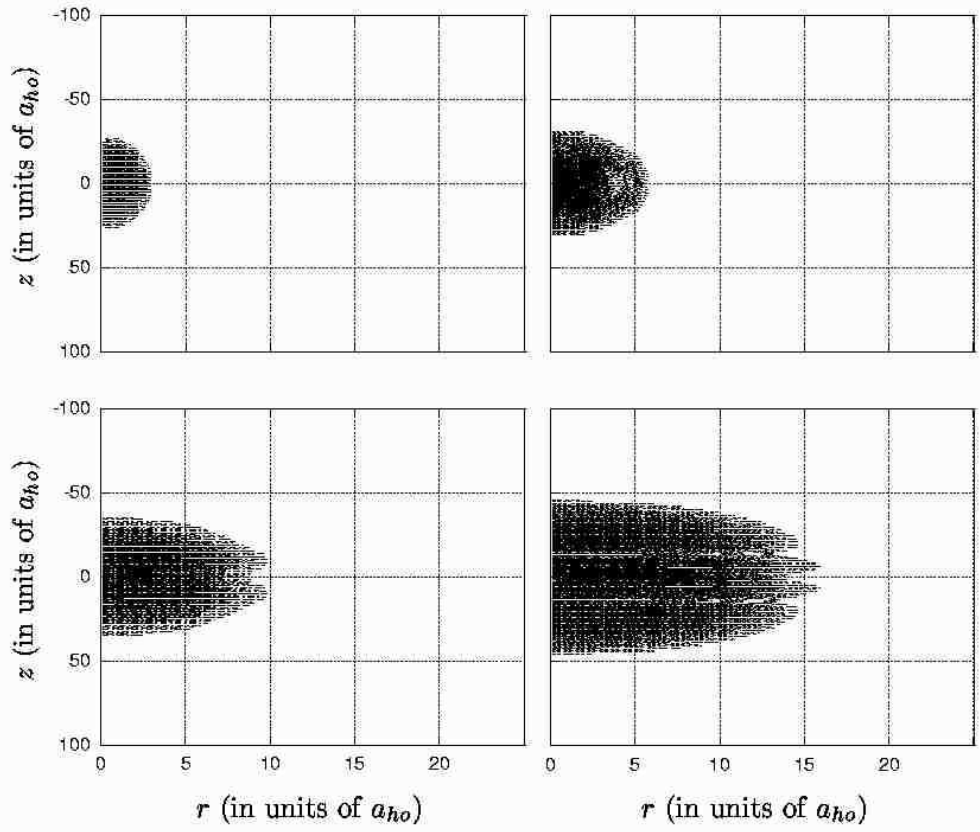,width=0.8\linewidth}}
\caption{
Expansion of a condensate interacting with its thermal cloud after release from
harmonic plus deep optical-lattice potential: 
snapshots of the
density distribution, shown as contour plots. From top left to bottom right:
$t=0$, $3.5$, $7$ and $11.5$ ms. The axial and radial coordinates are 
in units of $a_{ho}$.}
\label{expbecterm}
\end{figure}

\begin{figure}
\centerline{\epsfig{file=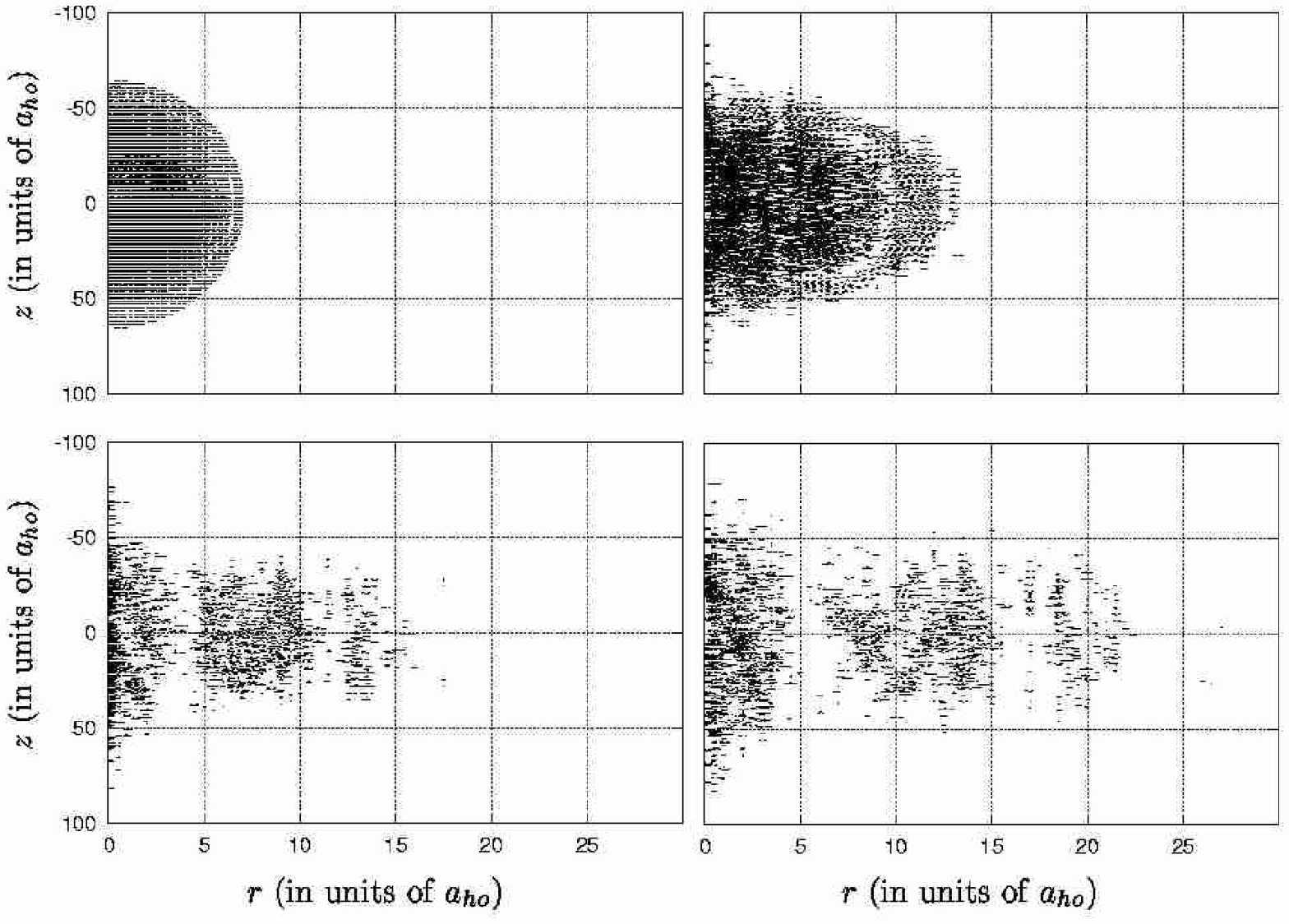,width=0.9\linewidth}}
\caption{
Expansion of a the bosonic thermal cloud interacting with a condensate 
after release from
harmonic plus deep optical-lattice potentail: 
snapshots of the
density distribution, shown as contour plots. From top left to bottom right:
$t=0$, $3.5$, $7$ and $11.5$ ms. The axial and radial coordinates are 
in units of $a_{ho}$.}
\label{exptermbec}
\end{figure}

We evolve the gas with a time-step $\omega\Delta t=7\cdot 10^{-5}$
after switching off both the harmonic trap and the periodic potential.
Snapshots of the atomic densities of the condensate and of its thermal
cloud, taken at the moment in which the 
potentials are switched off and
after  3.5, 7 and 10.5 ms of free expansion, are shown 
in figs.~\ref{expbecterm} and \ref{exptermbec}.
The condensate is seen in fig.~\ref{expbecterm} to develop side bands
which separate out of the central cloud, while the thermal cloud in
Fig.~\ref{exptermbec} simply spreads out.
These features of our numerical results reproduce those observed in the experiments
~\cite{pedri}.

The appearence of side bands in the condensate during expansion 
is due to Bragg scattering against the periodic
potential.
In fact, in a long-time simulation run of a one-dimensional model of the
expansion we have found that the condensate side-bands move at velocity
$v\simeq 2\hbar k_L/m$, corresponding to the momentum associated
with the first reciprocal vector of the optical lattice.
\section{Computational remarks}
\label{remark}
We have assessed the computational performance of the numerical method by repeating 
the test of Sec.~IV.A after changing either the
number of computational particles or the mesh size. 
We list in Table 1 the computational times 
elapsed while running the HPF-PGI-compiled code 
on a fully dedicated 1GHz Pentium III SCSI. 

These data provide the following values for the
specific CPU time costs of the GPE and VLE of the code per time-step:
\begin{equation}
\left\{\begin{array}{ll}
t_{GP} \sim 1.5 &\qquad\;\;\; \mu s/grid-point\\
t_{VL} \sim 0.6 &\qquad \;\;\; \mu s/particle\\
\end{array}\right.
\end{equation}
These figures invite a number of comments.
First, they show that the VLE section can evolve
just a few computational particles while
a single grid-point of the GPE solver is advanced. 
Since statistical accuracy requires of the order of $10-50$
particles per cell, we conclude that the VLE solver is
a potential computational bottleneck.

Let us nonetheless assume that the VLE and
GPE sections can evolve on a one-to-one CPU
time basis.
We can then focus on the grid
part only and estimate the feasibility of
large-scale applications
to finite-temperature condensates in
optical lattices.
Covering a simulation span of
$100$ $ms$ in steps of $0.1$ $\mu s$ requires
$10^7$ time-steps. At a cost of $1$ $\mu s$ per time-step
and grid-point, a grid with, say, $10^6$ grid-points
takes of the order of $10^6$ seconds, namely almost
two weeks of CPU time to complete. 

Ways to achieve substantial speed-up are clearly needed.
Among others, two promising (and not mutually exclusive)
strategies are non-uniform meshes and parallel
computing. Both strategies appear conceptually straightforward
and will be the object of future work.

%We see that reducing $NP$ by a 
%factor of two, shortens the computational time by $62\%$, while increasing 
%the mesh size by a factor of four increases the computational time by a 
%$22\%$. We thus see that it is crucial to find a compromise between the need 
%for lowering the statistical noise, and the need for keeping $NP$ small.   

\section{Concluding remarks}
\label{conclusions}
The increasing complexity and variety of phenomena observed in  
current studies of the dynamical behavior of 
normal and superfluid quantum 
gases at finite temperature motivate the development of 
suitable numerical tools to assist theoretical understanding. 

To this aim, we have combined a particle-in-cell method 
with an explicit time-marching algorithm to evaluate the
time evolution in models of a Bose-Einstein condensate and a cold-atom cloud.

We have tested the method against known analytical results  
in the simple physical situations offered by 
the expansion of a collisionless fermionic cloud 
without and with an inner Bose-condensed core.
We have also applied it to simulate novel experimental 
observations on the dynamical behavior  
of a condensate with its thermal cloud in a harmonic plus optical lattice 
potential, where we have found substantial accord with current experiments.

We have also analyzed those computational aspects
of the algorithm which are most relevant
to applications in large-scale
problems. This analysis emphasizes the need for
non-uniform meshes and parallel computing.
On the physics front, an
extension of the method 
to include the quantum collisional
integrals is under way.

\acknowledgments
This work was partially supported by INFM under PRA2001.
One of us (MLC) thanks Dr. F.~S. Cataliotti and 
Dr. S. Burger for making 
their experimental results available prior to publication.

\newpage

\newpage

\newpage

\newpage
\newpage

\newpage
\newpage
\newpage
\newpage
\newpage
\newpage
\newpage

\newpage

\begin{table}
\begin{tabular}{|c|c||c|}
\hline
P&$N_r\times N_z$& CPU-time (hh:mm:ss)\\
\hline
$1.6\cdot 10^6$&201$\times$401&8:35:27\\
\hline
$8\cdot 10^5$&201$\times$401&5:33:55\\
\hline
$1.6\cdot 10^6$&401$\times$801&10:22:24\\
\hline
\end{tabular}
\caption{
Expansion of
a cloud of fermionic $^{40}K$ atoms after release from the harmonic trap.
CPU-time (third column) elapsed on a 1GHz Pentium III SCSI for simulating 
$20000$ time-steps, corresponding to $20\; ms$, for various numbers of
computational particles (first column) and mesh sizes (second column).  }
\label{tavola}
\end{table}

\end{document}